# Idiosyncratic Approach to Visualize Degradation of Black Phosphorus

Bilal Abbas Naqvi [1,2], Muhammad Arslan Shehzad [1,2], Janghwan Cha,[2,3], Kyung Ah Min[2,3], M. Farooq Khan[2,3], Sajjad Hussain[1,2], Seo Yongho [1,2], Suklyun Hong [1,2], Eom Jonghwa [1,2], Jung Jongwan [1,2*].

[1]Departement of Nanotechnology &Advanced Materials Engineering, Sejong University, Neungdong-Ro 209, Seoul, 05006, South Korea

[2]Graphene Research Institute, Sejong University, Neungdong-Ro 209, Seoul, 05006,, South Korea.

[3]Departement of Physics & Astronomy, Sejong University, Neungdong-Ro 209, Seoul, 05006, South Korea.

Black Phosphorus (BP) is an excellent material for post graphene era due to its layer dependent band gap, high mobility and high $I_{on}/I_{off}$. However, its poor stability in ambient poses a great challenge in its practical and long-term usage. Optical visualization of oxidized BP is the key and foremost step for its successful passivation from the ambience. Here, we have done a systematic study of the oxidation of BP and developed a technique to optically identify the oxidation of BP using Liquid Crystal (LC). Interestingly we found that rapid oxidation of thin layers of BP makes them disappear and can be envisaged by using the alignment of LC. The molecular dynamics simulations also proved the preferential alignment of LC on oxidized BP. We believe that this simple technique will be effective in passivation efforts of BP and will enable it for exploitation of its properties in the field of electronics.

1. Introduction

2D layered material e.g. graphene and transition metal di-chalcogenides (TMDs) have opened a new era for electronic devices. In post graphene material, Black phosphorus (BP) stands out of them due to its unique electronic and structural properties. It has strong in-plane anisotropy due to a puckered orthorhombic crystal lattice in which one phosphorus atom (P) is bonded to three other P atoms. A monolayer of phosphorene has a direct band gap of 2 eV which decreases to 0.3 eV in bulk. The reason for this transition in band gap lies in the increase of layer to layer interaction in van der Waal structure.[1,2] Presence of band gap and high value of charge mobility make BP a promising candidate for field effect transistors with effective switching[2,3]. Due to its unique electronic structure, BP has also shown excellent performance in optoelectronics and memory devices.[4,5] BP shows both p and n-type behaviour as well as ambipolar transport characteristics depending upon the contact metal used.[3]

However, due to the presence of lone pairs with each phosphorus atom in BP, it is strongly reactive and readily oxidized. Oxidation is one of the most crucial problems in BP. The surface of relatively thinner BP flakes, which are the most suitable candidates for electronic devices, gets oxidized within an hour of cleaving from bulk if exposed to ambient environment.[1] This oxidation causes irreparable damages to the physical characteristic of BP .[6–8] Furthermore, long exposure leads to reacting with oxygen atoms in the ambient which cause etching of the surface and become invisible.[1]Till then this oxidation doesn't just confine to the surface but produce adverse effects such as volumetric changes, induced imbalanced dipole and formation of complex oxides species ($P_xO_y$) .[1,9–11] Till now AFM is the only topographic tool to assess the degree of oxidation by surface topography of oxidized BP.[12,13] TEM is also a powerful microscopic tool to study the oxidation of BP.[11,14] However, these techniques require dedicated and complex equipment as well as delicate sample preparation. Moreover, TEM has limitation to study only a tiny area at once and requires longer times for complete analysis.Until this report, there is no reported method of optically visualizing BP readily after its surface gets oxidize.

Previously, defects in 2D materials such as oxidation defects and grain boundaries are studied with Liquid Crystal (LC) alignment on them.[15,16] Defects such as oxidation defects and pin holes on a single domain of graphene were visualized using nematic LC.[17] LC has unique ability to orient themselves on different surfaces depending upon weak physical interaction which arise due to the electrical/magnetic field, surface modification and presence of defects on the surface.[18,19] LC are anisotropic materials and exhibits optical birefringence. 2D materials have broken pi stacking which favours the adsorption and alignment of LC on their surfaces. This alignment strongly depends upon crystalline orientation. This alignment is disturbed due to any surface modification or change in the crystalline orientation. Nematic phase of liquid crystal gained much significance after their alignment has been studied on synthetically grown 2D materials e.g. graphene, hBN and TMDs for optical visualization of their grains.[15,20,21] In nematic phase, all the molecules in liquid crystal stack parallel to an axis called director. Birefringence property of nematic LC help to optically analyze minute changes based on the alignment of LC. This birefringent behaviour in LC makes them suitable option to study anisotropy as well as surface defects. LC has been previously used to optically study grain boundaries and surface defects in graphene and TMDs.[17]

Here in this work, nematic LC (5CB) was used to optically study the oxidation defects on few layers BP flake under polarized optical microscopy (POM). UV light was used to do an in-depth analysis of oxidation by enhancing the oxidation of BP (details are in the methods section). LC alignment enables us to visualize the oxidation defects which were created on the surface either by ambient and UV light. We believe this work opens a room to not only visualize the defect of BP but also study the alignment behaviour of LC molecules on BP.

## 2. Results and Discussion

BP was exfoliated from bulk using micromechanical cleavage method and then transferred using PDMS stamp on Si/SiO$_2$ substrate (Figure 1c.). The gold pattern was fabricated to mark the coordinates of the flake. LC 5CB was coated on the sample using spin coater. Polyvinyl Alcohol (PVA) was coated on a thin glass slide and then rubbed unidirectionally. It was placed on the sample with PVA side contact with coated LC film. These grooves in thin PVA film helps in unidirectional alignment of LC molecules. In this way, only the alignment on desired alignment film (which, in this case, is BP) can be studied independently. (Schematic is shown in Figure 1b.)

In the Figure 1a,b the schematic representation of LC alignment on BP flake is shown. Due to broken pi stacking, LC has a tendency to align on the 2D materials. In BP, there are two primary atomic orientations, zigzag and armchair which are at 90º of each other. With molecular dynamics simulations, it turns out that LC tends to align preferentially along armchair orientation (details in molecular dynamics simulation section). Figure 1c. is the optical image of few layers BP flake which was transferred on Si/SiO$_2$ substrate using dry transfer technique. This flake was coated with LC and observed under POM. In Figure 1d. it can be seen that LC coated BP flakes exhibit a single color. A pristine BP flake is exfoliated from a single crystal and has single crystalline orientation throughout. Therefore, LC has aligned uniformly on this flake and this flake appears bright when seen with POM. In Figure 1d. the flake appears brighter which means that the director axis of aligned LC is in between the perpendicularly cross analyzer and polarizer. Figure 1e. is the Raman spectra of BP with and without LC coating. Three signatures peak $A_g^1$, $B_{2g}$ and $A_g^2$ are present in Raman spectra of pristine BP. While in LC coated BP among these three signature peaks of BP, peaks of C=C, C=N and biphenyl stretching peaks from LC are also present.

As discussed earlier few atomic layers thick BP referred as phosphorene is unstable in ambient conditions. BP has a strong puckered structure in which phosphorus (P) atoms are covalently bonded to 3 other phosphorus atoms. Each P atom has a lone pair which makes BP highly reactive in ambient.[10] Although this surface oxidation doesn't cause the lattice distortion, however, due to the formation of P-O bonds two strong dipoles are created rendering BP hydrophilic. This hydrophilicity is the further cause of more complex defect formation. This may cause volumetric changes in the lattice and eventually lead to complete loss of BP due to the formation of complex oxides and acidic species.[8,10]

To further extend the analysis of oxidation of BP and alignment behaviour of LC on BP, a triangularly shaped flake, shown in Figure 2a. was exposed to the atmosphere to get oxidized. This flake was a relatively thicker flake and keeping this in view it was exposed to ambient conditions for 48 hours. It gets oxidized consequently, which is evident in the Figure 2b. This flake was then coated with LC and observed under POM. Interference segments observed on this flake indicates that LC has several alignments on this flake. These different alignments indicate non-uniformity which evidently proves the presence of oxidations states on BP (Figure 2c.). When the sample was coated with LC and observed under POM, interestingly we found remnants of completely oxidized BP (Indicated as Pt. 1 in Figure 2b.), which was invisible otherwise. LC molecule aligns on these remnant oxides due to weak physical interaction. Due to birefringence of aligned LC molecules, this heavily oxidized flake can be seen. However, it is worth noting that in the absence of LC there is no visual sign of the presence of these residual oxides at Pt1. Thus LC serves as a powerful and simple ocular tool to visualize the obsolete BP and its remnants after oxidation.

Raman analysis was done at Pt.1 and Pt.2 (as indicated in Figure 2b.) in order to confirm the nature of these residual oxides (Figure 2d.). At Pt.2 three signature peaks of BP $A^1_g$, $B_{2g}$ and $A_g^2$ are present at 362 cm$^{-1}$, 439 cm$^{-1}$ and 468 cm$^{-1}$ respectively. Apart from these three modes, Raman analysis also shows small broad but prominent peak around 800-900 cm$^{-1}$. This broad peak corresponds to vibration modes of $P_xO_y$.[22,23] At the Pt. 1 with Raman analysis doesn't provide any evidence for BP. However, a relatively strong and broad peak around 900-1000 cm$^{-1}$. This peak corresponds to mixed signals from vibrations of phosphoric acids and also some of the vibrations of phosphoric oxides.[24] LC was coated on the sample and a Raman spectrum was also collected (indicated by a green star) at the oxidized region with LC coating. However, we find no considerable shifts in the newly emerged peak. There is a little decrease in the intensity of this peak when Raman spectra were taken with LC coated samples. This decrease in the intensity can be attributed to the fact that aligned LC molecules can suppress vibration coming from phosphoric acids and phosphoric oxide. The emergence of vibration modes from phosphoric acids is due to the formation of acidic species by the reaction between oxides and moisture present in ambient. Formation of acidic species leads to subsequent etching of BP making it almost invisible. Another BP flake was cleaved and subjected to $O_2$ plasma treatment for different. After 15 sec of plasma treatment the flakes is disappeared when observed in the optical microscope, however, LC coating makes it visible under POM (Figure S3).

2D materials e.g. and $MoS_2$ undergoes oxidational decay upon UV irradiation[15,17,25]. BP behave differently under dark and illuminated environments.[11] Thinner flakes of BP are more prone to oxidation and have faster oxidation kinetics. Monolayer BP has a direct band gap of 1.51 eV which favours oxidation as the redox potential of $O_2/O_2^-$ lies in this energy gap. Excitons generated from the conduction band of BP radicalized $O_2$ which reacts with BP. However, with an increase in a number of layers the band gaps increase (0.67 eV for bulk) and eventually fall below the redox potential of $O_2/O_2^-$ making oxidation hard for bulk BP.[26] However UV light can make oxidation thermodynamically favourable for Bulk BP by exciting excitons from the conduction band. Also, it can increase the rate of oxidation for thinner BP whose oxidation was already favourable without any UV exposure.

$$O_2 + \hbar\nu \longrightarrow O_2^- + h$$

Considering UV can make reaction kinetics of oxygen and BP fast, BP was subjected of UV exposure. To observe defects induced by UV on BP, a thin flake of was exposed to UV light in ambient for 2, 4 and 6 min. After each exposure samples were coated with LC and then observed under POM with cross-polarizer (Figure 3 b,c,d). The presence of defects can be clearly seen as shiny spots which emerge due to arise of new birefringence of LC on $P_xO_y$. Upon UV exposure, the surface of BP flake gets oxidized which leads to the change in LC alignment along new preferential orientation. Interestingly no oxidation streak was observed after the removal of LC Figure 3e. No defects are visible in this image and it is same as that of Figure 3a which is optical image of the freshly cleaved flake. However, when LC is coated on the same sample, oxidation defects are clearly seen as shiny spots. This experiment further strengthens our claim that LC can be used to investigate and visualize oxidation even in its earliest stage. It was also confirmed through DFT calculation that LC tends to align preferentially on oxidized BP as compared to pristine BP.

Raman spectra were also collected from pristine BP and after UV exposure for 2, 4 and 6 min to study the defect formation via phonon vibration mode. All the intensities of Raman active mode of BP show a decreasing trend with increasing UV exposure. (Figure 3 f,g). This analysis was done on two flakes with different thickness (see Figure S1) We found out that the decay in intensities is different for different thickness of flake and more sharp and prominent for the thinner ones. This may attribute to the faster kinetics of thinner flakes as the UV exposure intervals were kept constant for both samples.[8,9,11,26]

To gain further insight of BP oxidation and to study the surface of UV oxidized BP, AFM was employed. A freshly cleaved BP was subjected to AFM topographical analysis as shown in Figure 4a. The surface of pristine BP is relatively smoother and shows a height profile of 10 nm. This flake was exposed to UV exposure for 4 and 6 min (Figure 4b,c). With each UV exposure, the surface roughness of BP is increased. UV light increases the rate of oxidation of BP and surface roughness is the indication of ongoing oxidation (Figure 4b,c). It can be seen that there is a major change in the topology of BP. This change on the surface in-turn changes the alignment of BP making it obvious with visible light. As discussed earlier that oxidation increases the resistance of BP thus deteriorate its electrical characteristic. In order to study the effect of UV induced oxidation on electrical characteristics, a FET device was fabricated. A BP flake was transferred on PDMS stamp using micro-mechanical exfoliation using sticky tape and electrodes were patterned by e-beam lithography. 5nm of Cr and 50 nm of Au was thermally evaporated in a high vacuum chamber and deposited in the patterned electrodes. A BP FET was measured using standard 2 probe measurement. Figure 4d. shows the $I_{ds}$-$V_g$ curve for BP FET. After measuring the pristine sample the, the sample was subjected to UV exposure for 2, 4 and 6 min subsequently. It was noticed that there was a sharp decrease in the on current with an increase in UV exposure time (Figure 4d.). Charge carrier mobility was calculated for each curve (Figure 4e.) was calculated and we found a considerable decay in the charge carrier mobility with an increase in UV exposure time. Due to increase in oxidation, a number of charge trapping sites increases which hinders the mobility of charge carriers.

## 2.1 Molecular Dynamics Simulations

From our studies, it is affirmed that we can optically visualize BP and degradation (oxidation) on BP with LC. It was, therefore, imperative to further gain in-depth knowledge about LC alignment of BP and oxidized BP. To further extend our studies to the 5CB-BP system, Density functional theory (DFT) calculations are performed within generalized gradient approximation (GGA) for exchange-correlation (xc) functionals[27,28], implemented in the Vienna ab initio simulation package (VASP).[29,30] Figure 5a. is the calculated lattice of BP and the unit cell is indicated by A and B lattice parameters. Due to the fact that chemisorbed oxygen can form a number of bonds with BP, therefore; different BP-$O_2$ structures were considered (Figure S5). It was found that $P_4O_2$ was most stable energetically and was considered further for our calculations (Figure5b.). In this structure, one O atom is bonded to one P atom and due to large electronegativity difference between O and P (3.44 eV vs 2.09 eV respectively), P atom is being dragged into the lattice.[10] Due to this, lattice parameter along A and B direction varies to bond lengths in pristine BP with a considerable difference of 0.12 Å and 0.17 Å respectively (Figure 5 a,b). This variation could influence the anchoring of LC molecules on the surface.

A favourable stacking configuration can be determined by adsorption energy which is defined as $E_{ad.} = E_{5CB} + E_{sub} - E_{5CB-sub}$ , where $E_{5CB}$ and $E_{sub}$ is the energy of LC-5CB molecule and isolated the energy of BP substrate and $E_{5CB-sub}$ is the total energy of BP surface with 5CB molecules adsorbed on it. [15,31] First, LC preferential alignment was determined on pristine BP. Due to anisotropy in BP crystal lattice, there is only a unique possible stacking of BP along the zigzag direction (Figure 5c.). In this configuration one Bi-phenyl ring is attached in AB style while the other one is in A'A style. On contrary, along armchair atomic orientation 5CB molecule has the freedom to align in AB and AA stacking configuration. Based on calculated adsorption energies value, most favourable stacking mode is AB style along armchair direction. To compare preferential alignment of 5CB on BP and oxidized BP ($P_4O_2$), binding energies of 5CB on $P_4O_2$ was calculated. Our calculations show that out of all considered systems that LC alignment on oxidized Zig-Zag and Armchair direction is more stable as compared to bare phosphorous. Thus it further affirms our assertion that LC has a tendency to align on oxidized BP enabling us to visualize oxidation.

## 3. Conclusion

BP is highly unstable and readily oxidizes in ambient conditions. The kinetics of oxidation acts faster if it is exposed to UV light. The oxidation of BP can't be visualized optically; however, coating LC on oxidized BP helps in visualizing oxidation with POM. It was observed that birefringence property of LC can be utilized to study the oxidation behaviour of BP. Molecular dynamics simulation confirms that oxidized BP can have stronger binding with LC compared to bare BP which help to distinguish the oxidized and unadorned BP surface. This study will further help in improving the passivation techniques to make BP more stable and can be a practically used fabrication of the electronic device.

## 4. Experimental Details

**4.1 Fabrication of BP-LC Cell:** BP purchased from 2D semiconductors, was exfoliated using micromechanical cleavage technique using sticky tape and then transferred to PDMS stamp. This cleaved BP was then transferred to Si/$SiO_2$ substrate with an oxide thickness of 300 nm. These flakes were used for LC alignment and degradation studies. LC was coated using spin coater at

2500 RPM for 40 s to acquire film thickness of 0.2 µm. On the other hand, PVA was coated on a glass slide at 3000 RPM for 30 sec. This cover glass was then unidirectionally rubbed and placed on LC coated sample and samples. For UV light exposure homemade setup was used which was equipped with UV light wavelength of 220 nm and power UV lamp. After each UV light exposure samples were cleaned using acetone and methanol.

**4.2 Characterization:** To do alignment study on BP, samples were observed with Olympus BX-51 microscope fitted with rotatable polarizer and analyzer. Raman was done using Renishaw Raman apparatus with a laser wavelength of 514 nm and spot size of 0.1 µm. AFM was done in contact mode using Park system AFM system.

**4.3 Density Function Theory (DFT) calculations:** DFT calculations are performed within generalized gradient approximation (GGA) for exchange-correlation (xc) functionals, [27,28] implanted in the Vienna ab initio simulations package (VASP).[29,30] The kinetic energy cut-off is set to 400 eV, and electron-ion interactions are represented by the projector augmented wave (PAW) potentials.[32,33] Grimme's DFT-D3 method [34] based on semi-empirical GGA-type theory is used for Van der Waal correction. In the calculations, LC is absorbed on a (5x8) unit cell of monolayer BP and oxidized BP. For the Brillouin-zone integration, we use (lxlxl) grid for atomic optimization of absorbed 5CB on both BP and oxidized BP, in the Gamma centred scheme. Atomic coordinates are fully optimized until Hellman-Feyman forces are less than 0.01 eV/ Å.

**Supporting Information**

Additional Raman analysis, Plasma etching and LC alignment and additional results of the molecular dynamics simulation can be found in supporting info.

# Acknowledgement

H-Pu. *J. Chem. Phys.* **132,** (2010).

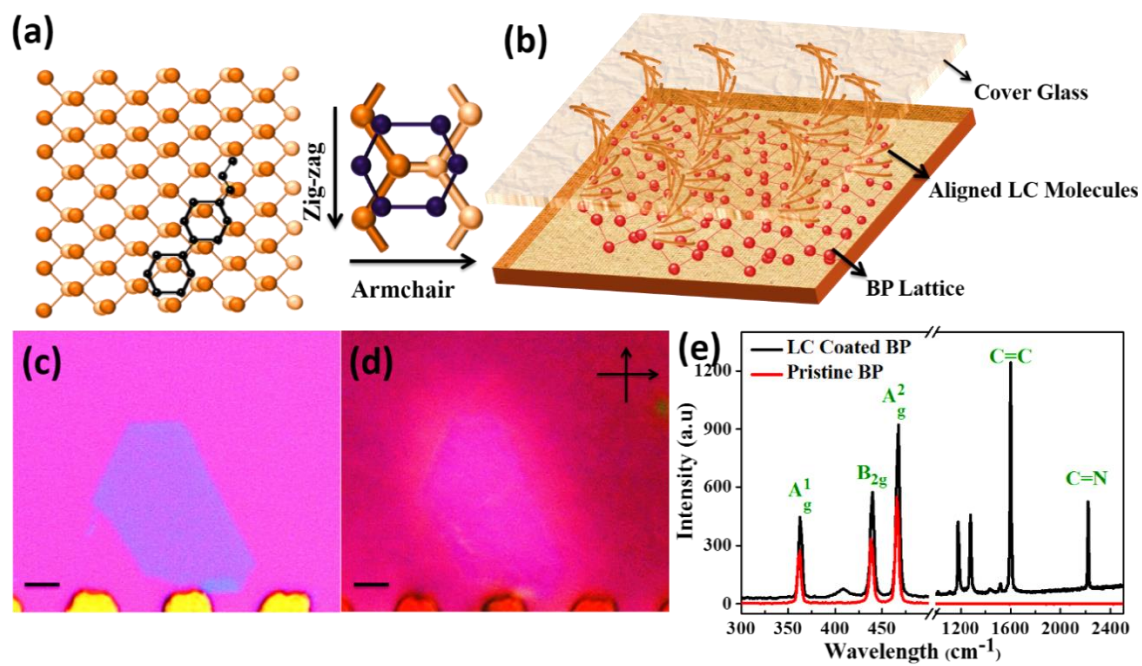

**Figure 1. LC alignment on BP.** (a) Schematic Top view of LC molecule on BP lattice**.** (b) Schematic of LC coated BP with a cover glass. (c) Optical image of BP Flake. (d) POM Image of

LC coated BP confirms alignment of LC (e) Raman spectra of Pristine BP and LC coated BP. (scale bar = 10 µm)

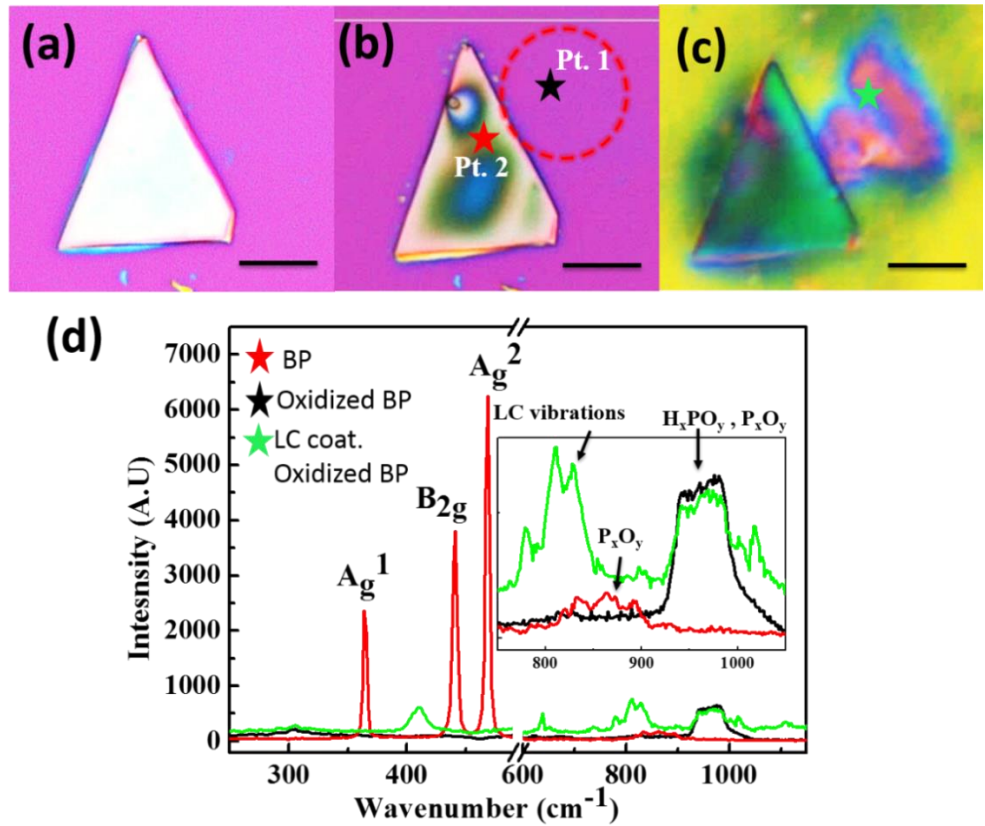

**Figure 2. LC alignment on oxidized BP due to ambient exposure and Raman analysis:** (a) Freshly Cleaved BP Flake (b) Oxidized BP Flake which was exposed to ambient; oxidation is visible. (c) LC coated BP in which oxidation remnants of a thin heavily oxidized flake are visible next to triangular flake. (d) Raman Spectra at three points indicated by stars. Red star indicates Raman taken on BP flake while black and green star indicates Raman at oxidizes region with and without LC. BP flakes show three characteristic sharp signature peaks which indicate a bulk flake. A small broad peak at 800-900 cm$^{-1}$ on BP corresponds to mixed phosphorus oxides vibrations. At the completely oxidized region, indicated by red and green star there is a relatively intense peak at 900-1000 cm$^{-1}$. This corresponds to mix signals from Phosphoric acids and Phosphorus oxides. The offset is zoomed of peaks emerging between 750-1100 cm$^{-1}$.

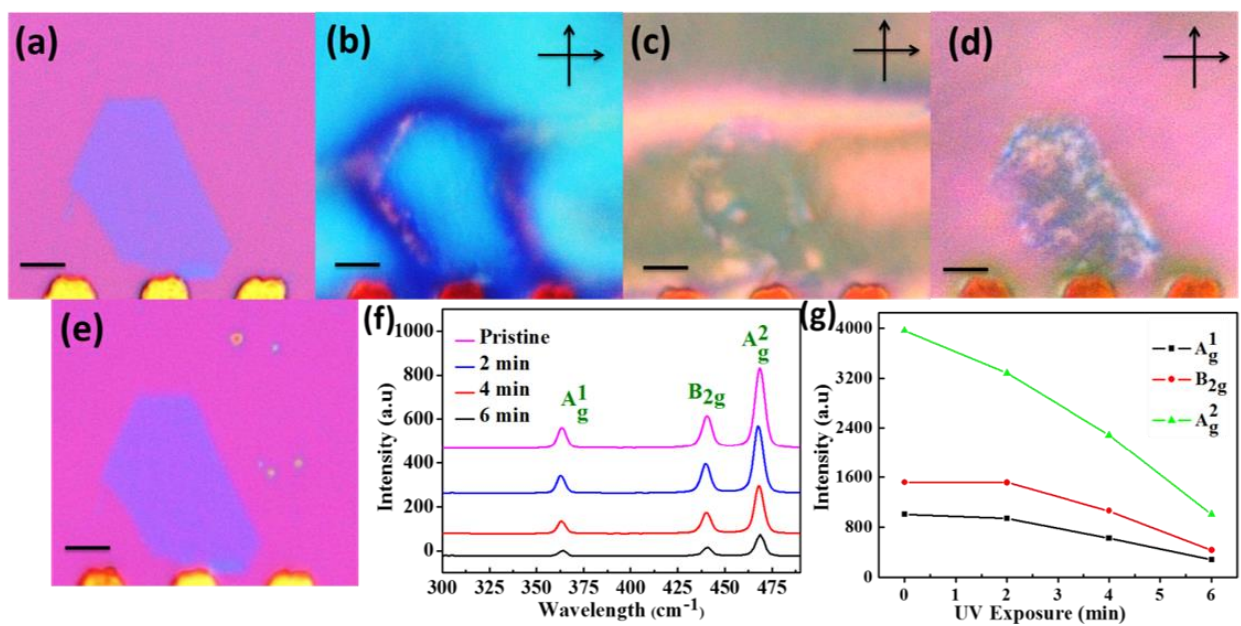

**Figure 3 LC alignment and Raman Analysis on UV Exposed BP Flake.** (a) Optical Image of BP before any exposure. (b,c,d) LC alignment on UV exposed flake for 2, 4 and 6 min. Increase in defect density is evident. (e) Optical image of BP flake after UV exposures. **(f,e)** Raman Spectra of BP Flake with subsequent UV exposure and Raman intensity plot respectively. (h) Raman Intensity plot of a relatively thicker flake. (scale bar = 5µm)

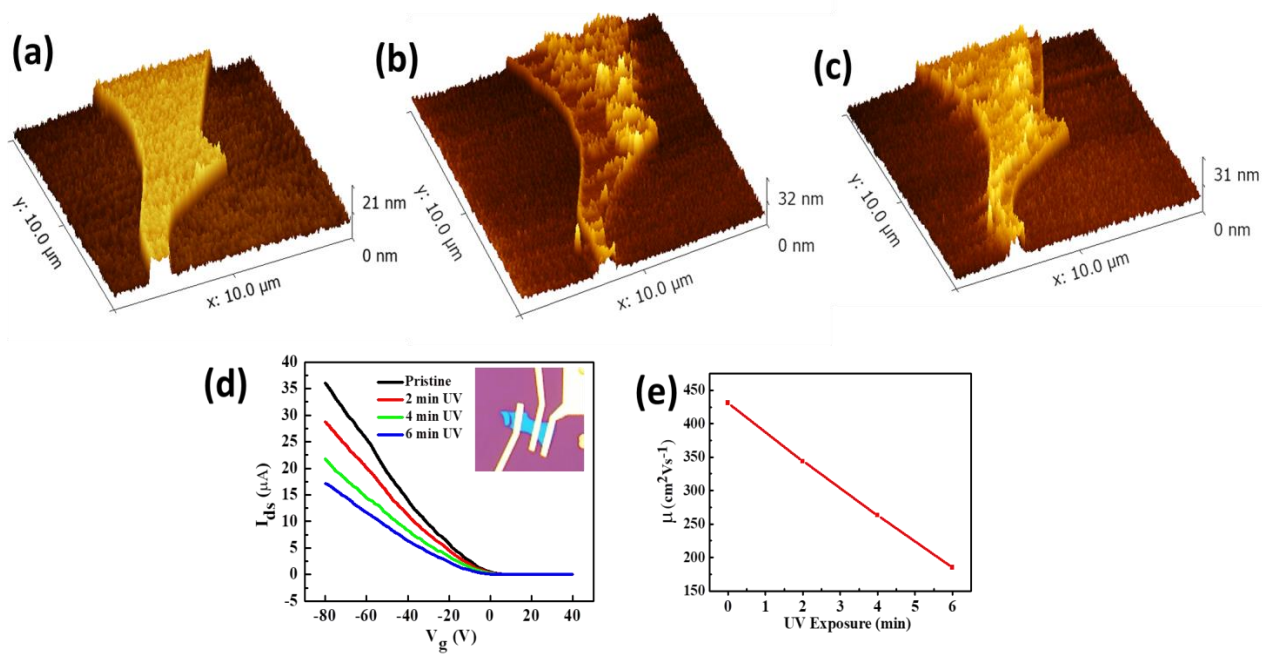

**Figure 4 AFM Topographical Analysis and electrical transport measurement with UV Exposure.** (a,b,c) AFM image of pristine BP, after 2 min UV exposure and min UV exposure respectively. Roughness has increased manifold with an increase in UV exposure. However, some points shows thinning which can be attributed to the formation of $H_xPO_y$. (d) The $I_{ds}$-$V_g$ curve of pristine and UV exposed BP. (e) Decreasing trend in mobility vs exposure time

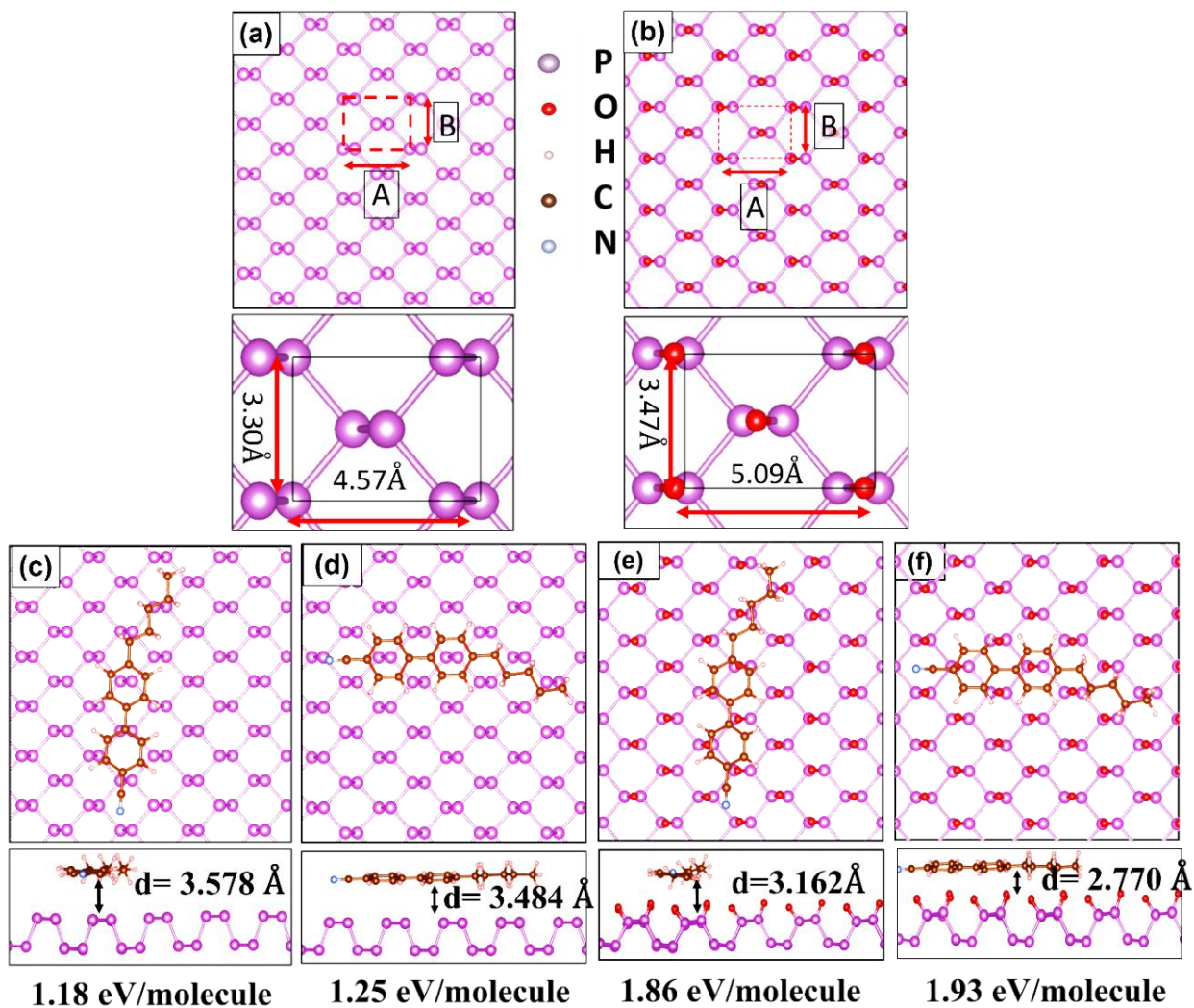

**Figure 5. DFT calculations result for possible stacking configuration of 5CB on BP and $P_4O_2$.** (a,b) top view of BP crystal lattice and $P_4O_2$. (c,d) 5CB molecule aligned parallel to zigzag and armchair orientation on BP respectively with adsorption energy and the distance between the centre of LC and BP lattice. (e,f) 5CB align along zigzag and armchair orientation on $P_4O_2$ respectively with adsorption energy and the distance between the centre of LC and BP lattice.